\documentstyle[aps,epsfig,prb,multicol, amsmath]{revtex}

\begin{document}

\preprint{{\it Accepted for publication in Applied Optics}}

\title{A simple method for the determination of slowly varying refractive index profiles from in situ spectrophotometric measurements}
\author{D. Poitras and L. Martinu\thanks{author to whom correspondence should be addressed; \protect \\  electronic-mail: lmartinu@mail.polymtl.ca}}
\address{Groupe de Recherche en Physique et Technologie des Couches Minces (GCM) and Department of Engineering Physics, \' Ecole Polytechnique, Box 6079, Station Centre-Ville, Montr\' eal (Qc), Canada H3C 3A7}
\date{\today}

\maketitle

\begin{abstract}
Reliable control of the deposition process of optical films and coatings frequently requires monitoring of the refractive index profile throughout the layer.  In the present work a simple in situ approach is proposed which uses a WKBJ matrix representation of the optical transfer function of a single thin film on a substrate.  Mathematical expressions are developed which represent the minima and maxima envelopes of the curves transmittance-vs-time and reflectance-vs-time.  The refractive index and extinction coefficient depth profiles of different films are calculated from simulated spectra as well as from experimental data obtained during PECVD of silicon-compound films.  Variation of the deposition rate with time is also evaluated from the position of the spectra extrema as a function of time.  The physical and mathematical limitations of the method are discussed.
\end{abstract}

\vspace{-8mm}
\begin{multicols}{2}

\section{Introduction}
In optical coatings, two different kinds of graded index profiles can occur:  (i) intentionally graded profile, when precisely controlled inhomogeneous optical coatings (for example, rugate filters) are produced;\cite{bovard93}  (ii) accidentally created inhomogeneities that may arise from the instability of the deposition process.  The latter may be due to different factors,\cite{ogura,sainty} including the change of microstructure of the film deposited, the contamination, the lack of control of the fabrication parameters, or the interaction of the growing film with the substrate.

In order to control the film quality it is, therefore, very important to calculate the refractive index depth profiles from in situ measurements.  Bovard,\cite{bovard85} inspired by the popular transmittance envelope method for the calculation of refractive index dispersion curves,\cite{manifacier,swanepoel,mouchart85} developed an approximative method based on the transmittance envelopes of the curve transmittance-vs-time.  In this work, we describe a generalized reflectance and transmittance envelope method based on WKBJ approximation.  A matrix representation of thin films with slowly varying index of refraction is used to obtain maxima and minima envelope expressions, which are then analytically solved and the refractive index profile is extracted.  We apply this method to evaluate the uniformity of the refractive index in amorphous hydrogenated silicon nitride (SiN$_{1.3}$) films deposited in low pressure plasma.

\section{Experimental apparatus}
\begin{figure}
\begin{center}
\epsfig{figure=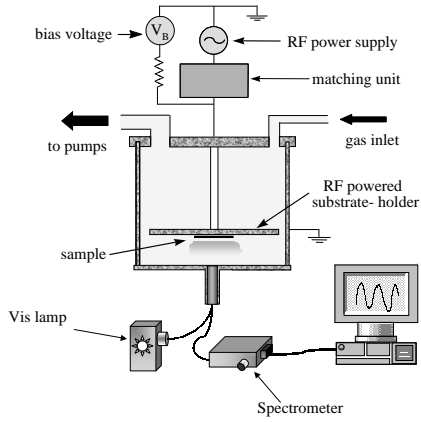,width=8.2cm}
\vglue0.1cm
\begin{minipage}{8.5cm}
\caption{Multiple wavelength optical monitor mounted on a plasma deposition chamber.\label{apparatus}}
\end{minipage}
\end{center}
\end{figure}

Before we present the theoretical background of our calculations, we first describe the experimental methodology for the preparation of optical films.  The films studied in this work were grown on a radiofrequency (13.56 MHz) -powered electrode (18 cm in diameter) in a plasma system (Fig.~\ref{apparatus}) described earlier.\cite{poitras}  A mixture of silane and ammonia (typically 1:3 ratio) was used for the deposition of SiN$_{1.3}$ on both glass and silicon wafer substrates, using a working pressure of 40 mTorr.  In order to ease the optical reflectance analysis, the back surface of the glass substrate was roughened.

\begin{figure}
\begin{center}
\epsfig{figure=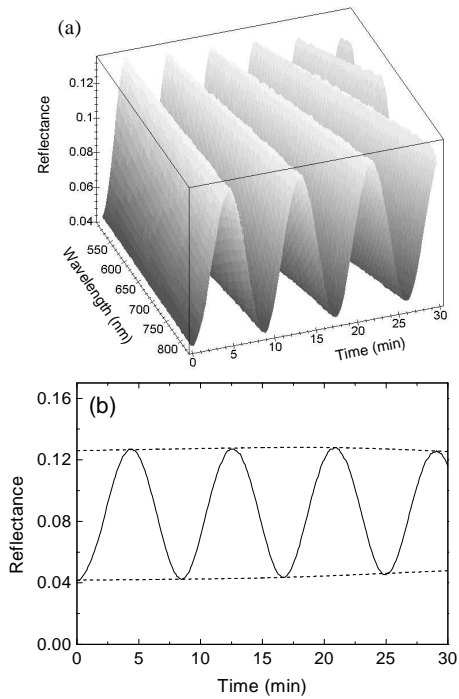,width=8.2cm}
\vglue0.1cm
\begin{minipage}{8.5cm}
\caption{(a) Typical spectra evolution measured with the optical monitor. (b) Closer look at the reflectance modulation with time at $\lambda$ = 789 nm. \label{Rt3d}}
\end{minipage}
\end{center}
\end{figure}

During the optical reflectance experiment, a white light beam (100 W halogen lamp, Oriel) hit the surface of the growing film at an angle of incidence close to the normal.  The reflected light was collected by an optical fiber and brought to a spectrometer (Multispec I, Oriel) equipped with a photodiode array (1024 diodes, Instaspec II, Oriel).  The acquisition time of the detector was kept between 0.3 s (Si substrate) and 0.6 s (glass substrate).  Light from the plasma was recorded as a background and subtracted from the spectra.  Each 4--6 seconds during the deposition, a spectrum was recorded.  Example of reflectance data measured in situ with the optical monitor is shown in Fig.~\ref{Rt3d}.  One can see the variation of the maxima and minima of reflectance both with time and wavelength.  A cross section of the spectrum in Fig.~\ref{Rt3d}(a) is shown in Fig.~\ref{Rt3d}(b), which represents the variation with time of the reflectance at a wavelength of 789~nm.

In addition, ex situ reflectance and spectroellipsometric measurements have been made using a Perkin-Elmer lambda-19 spectrophotometer and a variable angle spectroscopic ellipsometer (VASE, J. A. Woollam Co.), respectively.

\section{Description of the method}
\label{method}
The method used to determine the refractive index of the film at any time of the deposition process is based on the evaluation of the envelope curves connecting the reflectance minima and maxima (dashed lines in Fig.~\ref{Rt3d}(b)).  Following a WKBJ (Wentzel-Kramers-Brillouin-Jeffries) approximation,\cite{landau} one can express the ``optical transfer function'' of an inhomogeneous thin film by the following characteristic matrix:\cite{jacobsson}
\begin{equation} 
{\cal M} = \begin{bmatrix}
  {\sqrt {\dfrac {n_{in}}{n_{out}}}} \cos \delta 
                   & \dfrac{\mathrm{i}}{\sqrt{n_{in}\, n_{out}}} \sin \delta  \\ 
  {\mathrm{i}} {\sqrt{n_{in}\, n_{out}}} \sin \delta
                   & \sqrt{\dfrac{n_{out}}{n_{in}}} \cos \delta 
\end{bmatrix} 
\label{matrix}
\end{equation} with $\delta = (2\pi/ \lambda) \int_{0}^{d}[n(z)-{\mathrm{i}}k(z)]dz$ for normal incidence, where n$_{in}$ and n$_{out}$ are the film refractive index values near the film-substrate interface and near the surface, respectively, d is the thickness of the layer, z is the distance from the interface inside the layer, and n(z) is the refractive index profile in the layer.  From this characteristic matrix one can obtain expressions for the transmittance T and reflectance R of a film on a substrate:
\begin{equation}
T={\dfrac{n_s n_{out}}{n_0 n_{in}}}\left |{\frac {t_{in}t_{out} \exp (\delta)}{1+r_{out}\,\breve{r}_{in}\,\exp (2\delta)}}\right |^{2},
\label{T}
\end{equation}
\begin{equation}
R=\left |{\frac {r_{out}+\breve{r}_{in}\,\exp (2\delta)}{1+r_{out}\,\breve{r}_{in}\,\exp (2\delta)}}\right |^{2}.
\label{R}
\end{equation} Here, $t_{out}=2n_0/(n_{out}+n_0)$, $t_{in}=2n_{in}/(n_{in}+n_s)$, $r_{out} = (n_0-n_{out})/(n_0+n_{out})$ and $\breve{r}_{in} = (n_{in}- n_s+{\mathrm i} k_s)/(n_{in}+n_s-{\mathrm i} k_s)$.  The sign over $r_{in}$ indicates that this parameter can have both complex or real values, as one can consider the reflectance problem for both transparent and absorbing substrates.  Not surprisingly, the expressions above are very similar to those describing the homogeneous case, and they translate very well the fact that the WKBJ approximation is equivalent to taking into account only the reflections occurring at the interfaces in the coatings, while the particular shape of the refractive index inside the layer is neglected.

From the approximate expressions (\ref{T}) and (\ref{R}), one can obtain expressions for the minima and maxima of the transmittance or reflectance curves, T(z) or R(z).  From the data available (monitored R(z) or T(z) values) one can then evaluate experimentally the envelopes and solve the envelopes expressions (such as Eqs.~(\ref{Rminmax1}) and (\ref{Rminmax2}) , see below) for n(z).  However, several assumptions must be respected: (i) the value of the film extinction coefficient is low (k $\ll$ n), so that only the k appearing in the exponential terms is significant;  (ii) once a material is deposited, its refractive index value is not altered by the following deposition on top of it;  (iii) the substrate is considered semi-infinite, which means that the contribution to the reflectance from the back side is negligible.

\subsection{Refractive index profile}
\subsubsection{Transmittance}
Bovard\cite{bovard85} has developed a method for the transmittance which consists of solving the equations for the envelopes of the transmittance minima and maxima for n$_{in}$, n$_{out}$ and the absorbance parameter ${\cal A} = \exp [(-4\pi /\lambda) \int_0^d k(z)dz ]$.  In the following we show a generalized solution, for low or high index substrates, based on his previous work.\cite{bovard85}  For d~$=$~0, one obtains:

\begin{align}
\begin{split}
  n_{in}&=\left (N_t+\eta\sqrt{N_t^2-n_0^2 n_s^2}\right )^{1/2}, \\
  & \quad N_t=\dfrac{n_0^2+n_s^2}{2} -2 \epsilon n_0 n_s \left ( \dfrac{T_{max}-T_{min}} {T_{max}+T_{min}} \right ), \label{ninT}
\end{split}
\end{align} and for d~$>$~0:
\begin{align}
\begin{split}
n_{out}(z)&={\dfrac{2n_{in}n_s n_0 \epsilon}{n_s^2-n_{in}^2}} \left ({\dfrac{T_{max}-T_{min}}{T_{max}T_{min}}} \right ) \\
& \quad +n_0 \left [1+ {\dfrac{4n_{in}^2n_s^2}{(n_{in}^2-n_s^2)^2}} \left ({\dfrac{T_{max}-T_{min}}{T_{max}T_{min}}} \right )^2\right ]^{1/2}, \label{noutT}
\end{split} \\
{\cal A} (z)&={\dfrac{\epsilon}{r_{in}r_{out}}} {\dfrac{\sqrt{T_{max}/T_{min}}-1}{\sqrt{T_{max}/T_{min}}+1}}, \label{AT}
\end{align} with $\epsilon = -1$ for $n_{in}>|\breve n_s|$, $\epsilon = +1$ for $n_{in}<|\breve n_s|$, and $\eta = 1$ for $n_{in}^2>n_0 |\breve n_s|$, $\eta = -1$ for $n_{in}^2<n_0 |\breve n_s|$.

\subsubsection{Reflectance}
The solution to the inverse problem, i.e. finding the variation of the optical parameters with depth from Eq.~(\ref{R}), consists of solving the following minima and maxima envelopes relations for three unknowns r$_{in}$, r$_{out}$ and ${\cal A}$:
\begin{align}
R_{min}={\frac {r_{out}^2-2\epsilon r_{out}\Re (\breve r_{in}) {\cal A} + |\breve r_{in}|^2 {\cal A} ^2 }{1-2\epsilon r_{out}\Re (\breve r_{in}) {\cal A} +r_{out}^2 |\breve r_{in}|^2 {\cal A} ^2 }}, \label{Rminmax1} \\
R_{max}={\frac {r_{out}^2+2\epsilon r_{out}\Re (\breve r_{in}) {\cal A} + |\breve r_{in}|^2 {\cal A} ^2 }{1+2\epsilon r_{out}\Re (\breve r_{in}) {\cal A} +r_{out}^2 |\breve r_{in}|^2 {\cal A} ^2 }}.
\label{Rminmax2}
\end{align}  Here, $\Re (\breve r_{in})$ is the real part of $\breve r_{in}$.  The system of Eqs.~(\ref{Rminmax1}) and (\ref{Rminmax2}) offers only two equations.  Therefore, to solve it we assume that at the first stage of the deposition (d~$\rightarrow$~0), the film is homogeneous (n$_{out}$~$ \rightarrow$~n$_{in}$) and transparent (${\cal A} \rightarrow 1$).  Using this approximation, and solving for n$_{in}$, we find

\begin{align}
n_{in} &= \sqrt{n_0 n_s} \left ( N_r + \eta \sqrt{N_r^2-{\dfrac{n_s^2+k_s^2}{n_s^2}}}\right ) ^{1/2},\label{nin}
\end{align}where \begin{align} N_r={\dfrac{(\epsilon +1)R_{min}-(\epsilon -1)R_{max}+2}{(\epsilon -1)R_{max}-(\epsilon +1)R_{min}+2}} \notag.
\end{align}

Keeping the value of n$_{in}$ constant, and solving for r$_{out}$ and for ${\cal A}$ when d~$>$~0, the calculated solutions are as follows:

\end{multicols}

\begin{align}
r_{out}(z)&=-\left [ -\dfrac{1}{2}(\dfrac{B}{A} -\zeta C)-\dfrac{1}{2} \sqrt{-4+(\dfrac{B}{A} -\zeta C)^2}\right ]^{1/2} , \label{rout} \\
{\cal A}(z)&=1/2\,{\dfrac {(1-{r_{out}^{4}}) (R_{max}-R_{min})} {\epsilon r_{out} \Re (\breve{r}_{in})\left [{r_{out}^{2}}(2R_{min} R_{max}-R_{max}-R_{min})+(2-R_{min}-R_{max})\right ]}} ; \label{AR}
\end{align} where
\begin{align*}
A&=|\breve{r}_{in}|^2 (R_{max}-R_{min})^2 , \\ 
B&=2\Re (\breve{r}_{in})^2 (2-R_{max}-R_{min}) (2R_{min}R_{max}-R_{min}-R_{max}) , \\
C&=\sqrt{2+\frac{B^2}{A^2}-\frac{D}{A}} , \\
\begin{split}
D&=-2 |\breve r_{in}|^2 (R_{max}-R_{min})^2+8\Re (\breve r_{in})^2 \left [ (R_{max}+R_{min})^2 +2 (1-R_{max}-R_{min}) \right.\\
&\left. +2 R_{max}^2R_{min}^2\left ( \dfrac{1}{R_{min}} +\dfrac{1}{R_{max}}-1\right ) \right ] ,
\end{split}
\end{align*} and
\begin{equation*}
\zeta =
  \begin{cases}
   +1& \text{if  } {\dfrac{1+r_{out}^4}{r_{out}^2}}>{\dfrac{B}{A}}, \\
   \\
   -1& \text{if  } {\dfrac{1+r_{out}^4}{r_{out}^2}}<{\dfrac{B}{A}}.
  \end{cases}
\end{equation*} \begin{multicols}{2} The refractive index profile n$_{out}$(z) is then calculated from $n_{out}=n_0 (1-r_{out})/(1+r_{out})$.

\begin{figure}
\begin{center}
\epsfig{figure=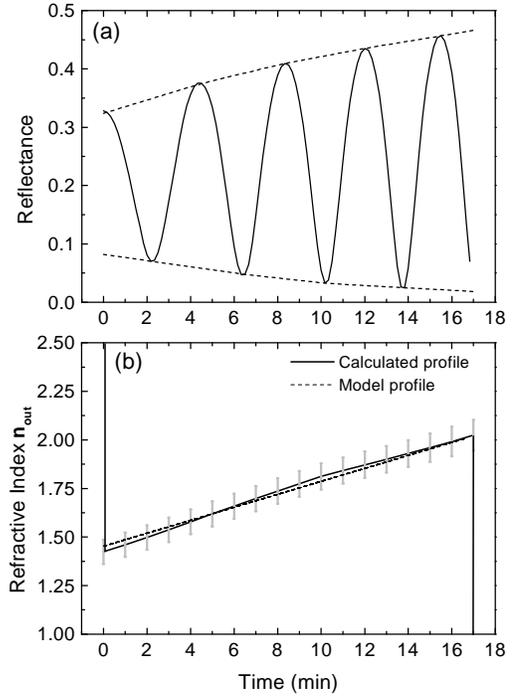,width=8.2cm}
\vglue0.1cm
\begin{minipage}{8.5cm}
\caption{Linear refractive index profile: (a) Reflectance vs time, (b) model (dashed line) and calculated (solid line) n(t) profiles ($\lambda = 800$ nm).\label{linear1}}
\end{minipage}
\end{center}
\end{figure}

To illustrate the method with an example, we applied the above expressions to the envelopes surrounding the simulated spectra of a 1-$\mu$m-thick transparent (k~$=$~0) layer with a linear profile and a deposition rate of 60~nm/min (Fig.~\ref{linear1}).  The profiles are expressed as a function of time instead of distance z, to simulate the type of data obtained experimentally with an optical monitor.  The calculated profile in Fig.~\ref{linear1}(b), although clearly approximate, is in very good agreement with the model linear profile used in the simulation.

\subsection{Extinction coefficient profile}
Calculation of k(z) from ${\cal A}$(z) in Eq.~(\ref{AT}) or (\ref{AR}) is done using the following relation:
\begin{equation}
k(z)= {\dfrac{-\lambda}{4\pi}}{\dfrac{d}{dz}}\ln {\cal A} (z).
\end{equation}

\begin{figure}
\begin{center}
\epsfig{figure=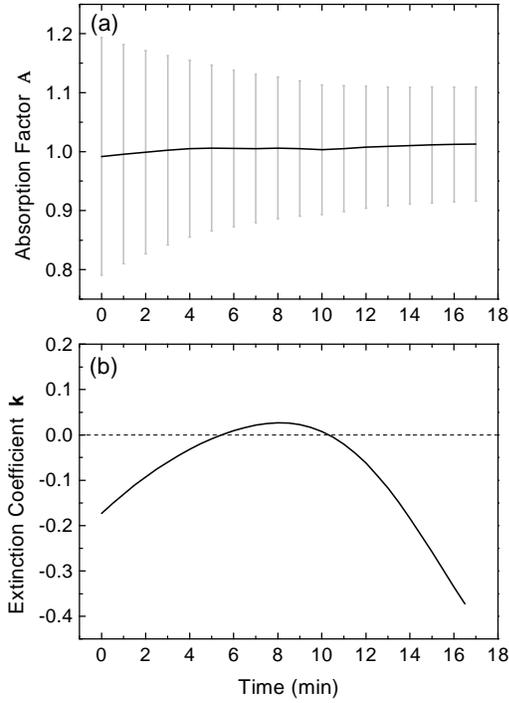,width=8.2cm}
\vglue0.1cm
\begin{minipage}{8.5cm}
\caption{Linear refractive index profile: (a) $\cal{A}$(t) profile, and (b) k(t) profile (dashed line: model, solid line: calculated).\label{linear2}}
\end{minipage}
\end{center}
\end{figure}

Figure~\ref{linear2} shows the ${\cal A}$(z) profile obtained from the envelopes of Fig.~\ref{linear1}(a).  The corresponding extinction coefficient profile is shown in Fig.~\ref{linear2}(b).  One can see that small deviations of ${\cal A}$(z) from 1 (the value it should have) have strong impact on the value of the calculated extinction coefficient, which may reach unexpected negative values.  Therefore, experimental precision of the envelopes must be very high if one is to find reliable k values.

\subsection{Deposition rate and physical position in the layer}
\begin{figure}
\begin{center}
\epsfig{figure=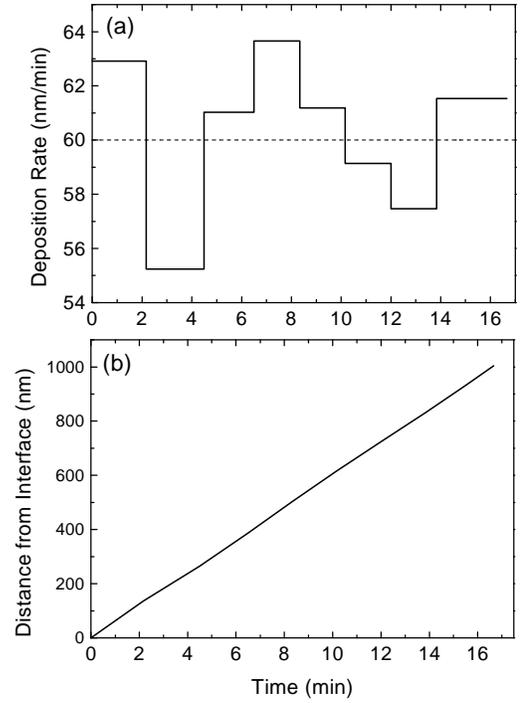,width=8.2cm}
\vglue0.1cm
\begin{minipage}{8.5cm}
\caption{Linear refractive index profile: (a) deposition rate values between successive extrema, (b) physical position in the layer as a function of deposition time (dashed line: model, solid line: calculated).\label{linear3}}
\end{minipage}
\end{center}
\end{figure}

It is interesting to express the results in Figs.~\ref{linear1} and \ref{linear2} as a function of physical position inside the layer instead of deposition time.  When the deposition rate is constant during the deposition, the transformation of the results as a function of thickness is obvious.  Unfortunately, it may not be the case for numerous experiments.  If the deposition rate varies during the deposition, one can still use the fact that the optical thickness between two successive extrema is equal to a quarterwave.  Using the relation $m\lambda/4 = \int_0^d n(z)dz$ (m is an integer representing the interference order) and assuming that the deposition rate is constant between successive extrema, one obtains the following relation:
\begin{equation}
{\frac{\lambda}{4}}\approx {\dfrac{z_2-z_1}{t_2-t_1}} \int_{t_1}^{t_2} n(t) dt, \label{zaxis}
\end{equation} where z$_1$, z$_2$ and t$_1$, t$_2$ correspond to the positions of two successive extrema on the thickness (distance) and time scales, respectively.  The relationship between the time and distance scales can be obtained from t$_1$~=~0 (z$_1$~=~0), where the first extremum occurs.  Figure~\ref{linear3}(a) shows the calculated evolution of the deposition rate from the simulation in Fig.~\ref{linear1}.  The deposition rate varies around the expected value (60 nm/min) within an interval of plus or minus 4 nm/min.  One can hence calculate the relation between the deposition time axis and the distance in the layer (see Fig.~\ref{linear3}(b)).  It appears that variation of the calculated deposition rate around its overall mean value is negligible and that total thickness found is consistent with the expected value of 1 $\mu$m.

\subsection{Error calculus}
\subsubsection{Transmittance}
A good estimate of the error in determining the n(z) and $\cal{A}$(z) values may be obtained from the calculation of the derivatives of Eqs.~(\ref{ninT})--(\ref{AT}).  In the case of the transmittance, assuming that n$_s$ is known precisely and that the error of T$_{max}$ and T$_{min}$ are the same and not related, we have

\end{multicols}

\begin{equation}
{\dfrac{\Delta n_{in}}{n_{in}}} \approx \left |{\dfrac{(n_{in}^2-n_0^2)(n_s^2-n_{in}^2)}{2(n_0n_s+n_{in}^2)(n_0n_s-n_{in}^2)}}\right | \left ( {\dfrac{T_{max}+T_{min}}{T_{max}-T_{min}}}\right ) {\dfrac{\Delta T}{T}},
\end{equation}
\begin{equation}
  \begin{split}
    {\frac{\Delta n_{out}}{n_{out}}} &\approx ({\frac{T_{max}+T_{min}}{T_{max}-T_{min}}}) {\frac{\Delta T}{T}}
   + {\frac{n_0C_n}{\sqrt{1+C_n^2}}}\left [\left| {\frac{n_{in}^2+n_s^2}{n_{in}^2-n_s^2}}\right | {\frac{\Delta n_{in}}{n_{in}}}\right ]
  \end{split}
\end{equation} with 
\begin{equation}
C_n=\left |{\dfrac{2n_{in}n_s}{n_{in}^2-n_s^2}} \right | \left ({\dfrac {T_{max}-T_{min}}{T_{max}T_{min}}}\right ).
\end{equation}Similarly:
\begin{equation}
  \begin{split}
    {\frac{\Delta {\cal A}}{{\cal A}}} &\approx {\dfrac{2\sqrt{T_{max}T_{min}}}{T_{max}-T_{min}}} {\frac{\Delta T}{T}} + {\dfrac{2n_0n_{out}}{n_{out}^2-n_0^2}} {\frac{\Delta n_{out}}{n_{out}}} + \left | {\dfrac{2n_sn_{in}}{n_{in}^2-n_s^2}}\right | {\frac{\Delta n_{in}}{n_{in}}}.
  \end{split}
\end{equation}

\begin{multicols}{2}

\subsubsection{Reflectance}
Expressions for the error in the envelope reflectance method can be found from the derivation of Eqs.~(\ref{nin})--(\ref{AR}) as functions of R and n$_{in}$.  The resulting equations are not shown here, but have been used for the determination of the error bars in the figures appearing in this work (see Figs.~\ref{linear1}, \ref{linear2}, \ref{exp1a}, \ref{exp2} and \ref{validity}).  The error on R$_{max}$ and R$_{min}$ have been fixed to a value of $\Delta$R~$=$~1\% and corresponds to the experimental error on the measured extrema.  This value doesn't take into account the error generated during the calculation of the envelopes.

\section{Application to experimental data}
\label{application}
\begin{figure}
\begin{center}
\epsfig{figure=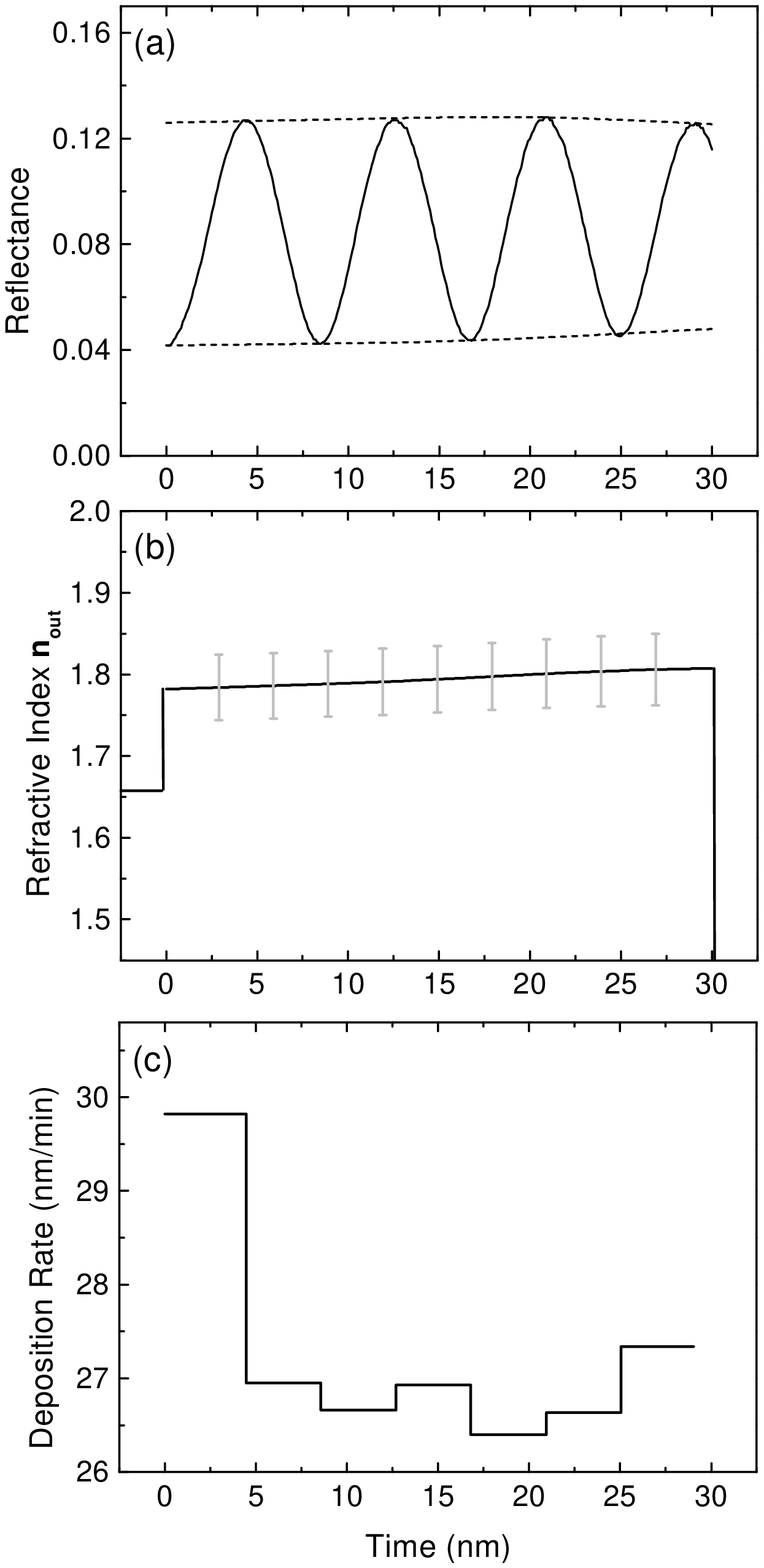,width=8.2cm}
\vglue0.1cm
\begin{minipage}{8.5cm}
\caption{Experimental data for an inhomogeneous SiN$_{1.3}$ film on glass: (a) Reflectance evolution, (b) calculated n(z) profile, (c) calculated deposition rate variation.($\lambda = 789$ nm)\label{exp1a}}
\end{minipage}
\end{center}
\end{figure}

\begin{figure}
\begin{center}
\epsfig{figure=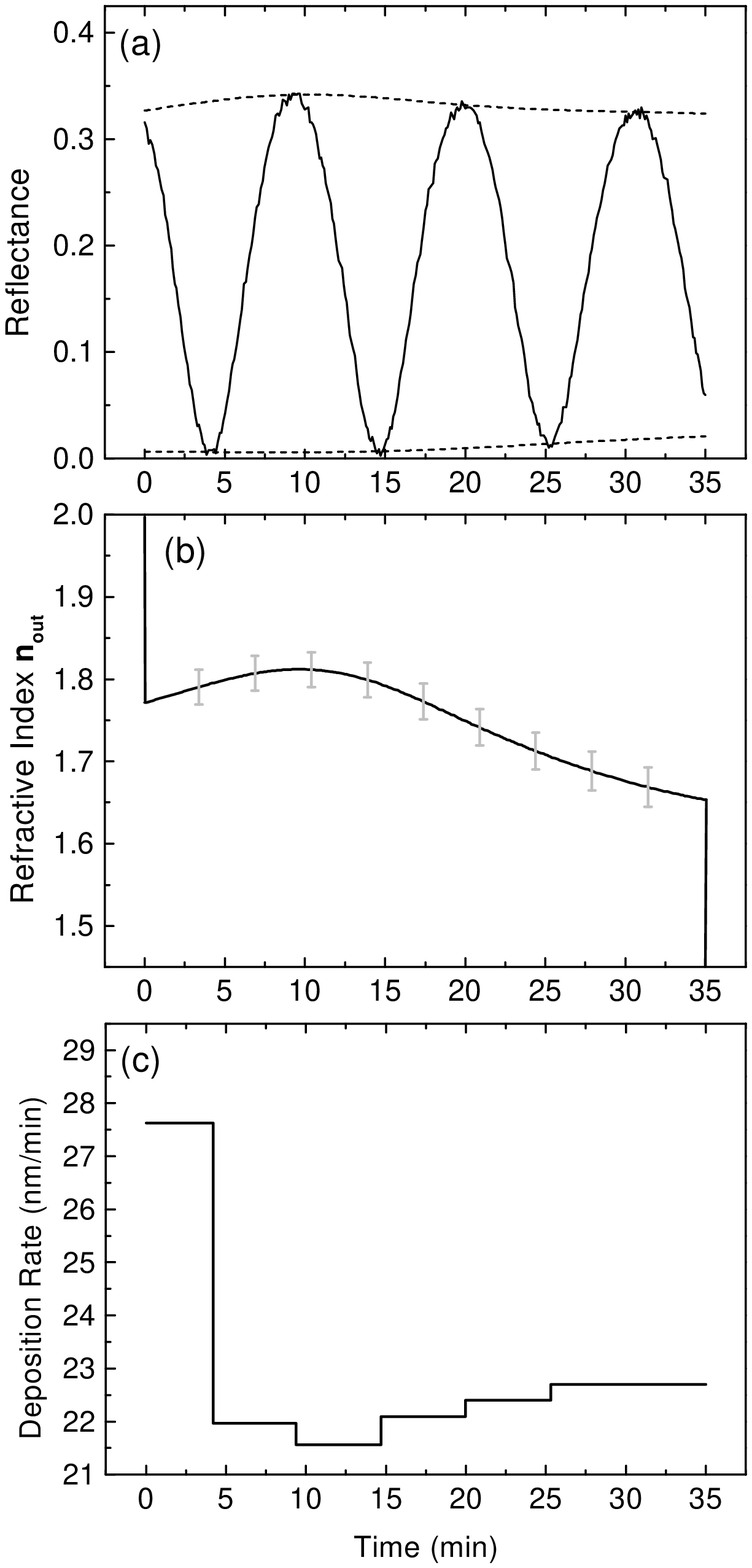,width=8.2cm}
\vglue0.1cm
\begin{minipage}{8.5cm}
\caption{Experimental data for an inhomogeneous SiN$_{1.3}$ film on silicon: (a) Reflectance evolution, (b) calculated n(z) profile, (c) calculated deposition rate variation.($\lambda = 826$ nm)\label{exp2}}
\end{minipage}
\end{center}
\end{figure}

The method has been tested experimentally for silicon nitride films deposited on both transparent (glass) and absorbing (silicon) substrates.  The results are shown, respectively, in Figs.~\ref{exp1a} and \ref{exp2}.  In these experiments, the deposition conditions were deliberately not precisely controlled: as a result, a refractive index gradient in the layers is observed.  The profile in Fig.~\ref{exp1a} is realistic; The variation in n(z) appearing in Fig.~\ref{exp1a} may be mostly due to oxygen contamination by water vapor present in the earlier stage of the film growth.  Additionally, n(z) in Fig.~\ref{exp2} may vary because of unstable deposition parameters, such as oxygen leak.  Index variations could also be due to changes in the microstructure of the layer (varying porosity).  Figures~\ref{exp1a}(c) and \ref{exp2}(c) show a higher deposition rate value in the initial stage of the deposition process;  further investigations are on the way to identify further experimental evidence of these phenomena.

\begin{figure}
\begin{center}
\epsfig{figure=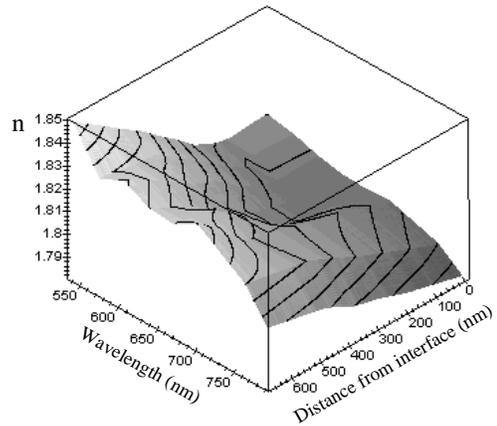,width=8.2cm}
\vglue0.1cm
\begin{minipage}{8.5cm}
\caption{Experimental data for an inhomogeneous SiN$_{1.3}$ film on glass: Evolution of the refractive index dispersion within the layer (same film as in Fig.~\ref{exp1a}).\label{exp1b}}
\end{minipage}
\end{center}
\end{figure}

Repeating the calculation for several wavelengths, the value of the refractive index dispersion at different positions in the layer can be obtained, as shown in Fig.~\ref{exp1b}.  The distance has been calculated from Eq.~(\ref{zaxis}); the mean value of the deposition rate has been used (22~nm/min).

\section{Discussion}
\subsection{Experimental error}
One must be very careful when using an optical monitor for the quasi-continuous measurement of transmittance or reflectance with time in a plasma environment.  For measurements at a single wavelength in a glow discharge, use of a lock-in detection system would help to subtract the light generated by plasma.  In our case, with a photodiode array, such light chopping was not used.  In addition, it was found that the main source of experimental error was the reflection and the scattering of light on different metallic components in the deposition chamber. Those were suppressed by installing a diaphragm between the optical monitor and the sample, and by blackening the surface of the sample holder.  Such perturbations will hardly change the phase information of the signal, but they can have a significant effect on the signal amplitude.

\subsection{WKBJ approximation limitations}
\begin{figure}
\begin{center}
\epsfig{figure=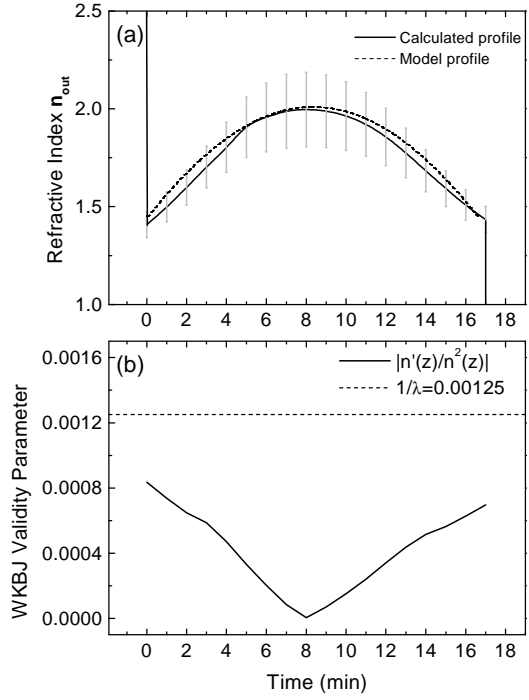,width=8.2cm}
\vglue0.1cm
\begin{minipage}{8.5cm}
\caption{Half-period-sinus refractive index profile: (a) model (dashed line) and calculated (solid line) n(z) profiles, (b) WKBJ validity condition from Eq.~(\ref{valid}) (dashed line: $1/\lambda$, $\lambda = 800$ nm).\label{validity}}
\end{minipage}
\end{center}
\end{figure}

As pointed out in Sec.~\ref{method}, the matrix representation in Eq.~(\ref{matrix}) follows the WKBJ approximation, which holds when 
\begin{equation}
\left | {\dfrac{1}{n^2(z)}} \nabla n(z) \right | \ll {\dfrac{1}{\lambda}}.
\label{valid}
\end{equation} The method has been tested on simulated envelopes obtained from a slightly more complex refractive index profile.  Figure~\ref{validity}(a) shows the model profile used and the calculated profile, while Fig.~\ref{validity}(b) shows an evaluation of the left (solid line) and right (dashed line) part of Eq.~(\ref{valid}).  One can see that the relation (\ref{valid}) holds throughout the profile.  Consequently, the WKBJ approximation is valid for these profiles, like it is for the other less complex linear profiles studied in this work.

\subsection{Envelopes-related limitations}
Precise determination of the envelopes from experimental spectra is delicate.  Algorithms for finding envelopes were developed and are used for the determination of refractive index dispersion from ex situ spectrophotometric measurements.\cite{minkov93,macclain}  In these cases, the spectra and its envelopes are functions of wavelength instead of time or distance, so that limit conditions at the right and left sides of the spectra can be used in order to improve the accuracy of the envelope computed.  No such limit conditions exist in our case.  Consequently, simple cubic or linear spline has been chosen to compute the envelopes.  As a result, the left and right extremes of the spectra are more likely to generate errors in the calculus.
In addition, to increase the accuracy of the envelopes, the layer must reach a minimum thickness, so that the spectra show at least 3 extrema.  The value of this minimum thickness will depend on the index of the film and substrate materials used, and on the wavelength of the probing light.  The shorter the wavelength, the smaller the minimum thickness will be.  Once such minimum thickness is reached, the method can be used to follow, in real time, the evolution of the refractive index profile during the film growth.

\subsection{Change of sign of $\zeta$ within the layer}
\begin{figure}
\begin{center}
\epsfig{figure=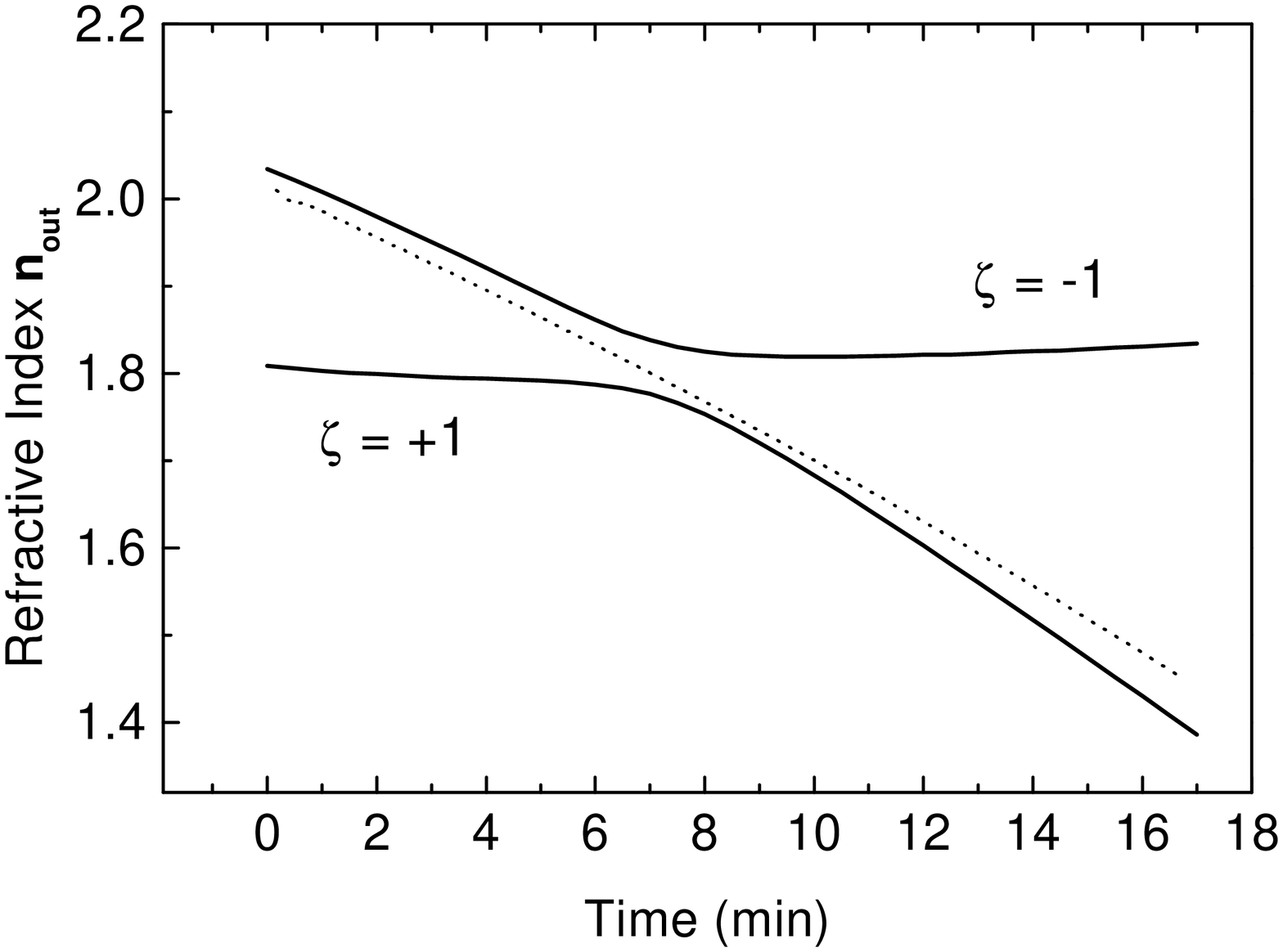,width=8.2cm}
\vglue0.1cm
\begin{minipage}{8.5cm}
\caption{Effect of of $\zeta$ on n(t) profiles (dashed line: model profile, solid: calculated profiles, $\lambda = 800$ nm).\label{zeta}}
\end{minipage}
\end{center}
\end{figure}

In Sec.~\ref{method}, the sign parameters $\epsilon$, $\eta$ and $\zeta$ can all reach values of +1 or -1, depending on the different values of the refractive index of the substrate and of the film.  The former two are easy to determine, as they depend only on the values of n$_s$ and n$_{in}$, and do not depend on z.  The latter one ($\zeta$) can be more difficult to determine since its value depends on n$_{out}$(z) and is changing along the profile.  Figure~\ref{zeta} illustrates the effect of a change in the sign of $\zeta$ on the refractive index profile: Similarly to Fig.~\ref{linear1}, a linear profile (dashed line) has been used to simulate the variation with time of the reflectance during the deposition of the graded layer.  The envelope method developed above has been used to calculate the refractive index profile twice, i.e. with $\zeta$ value of +1 and -1, respectively.  Around t~$\approx$~7~min., the variation of r$_{out}$, R$_{min}$ and R$_{max}$ is such that the value of $\zeta$ in Eq.~(\ref{rout}) should be changed from -1 to +1 in order to calculate an accurate refractive index profile.

\subsection{Ex situ analysis}
\begin{figure}
\begin{center}
\epsfig{figure=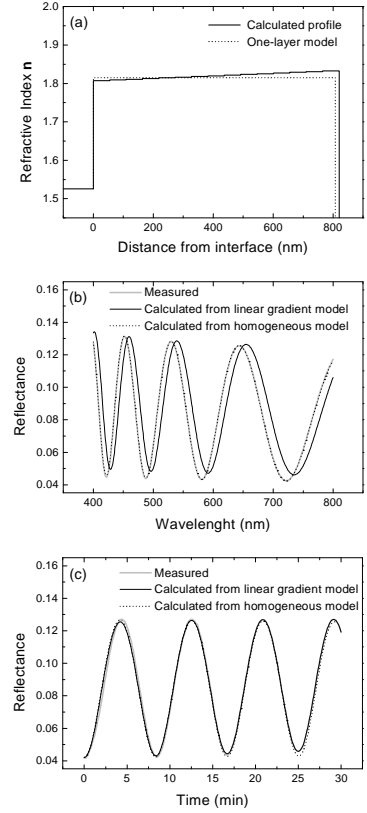,width=8.5cm}
\vglue0.1cm
\begin{minipage}{8.5cm}
\caption{Ex situ optical analysis of SiN$_{1.3}$ on glass (same sample as in Fig.~\ref{exp1a}): (a) n(z) profile calculated using the envelope method (solid line, same as Fig.~\ref{exp1a}(b)) and homogeneous one-layer model (dotted line) used to fit the ex situ measurements (in (b)); (b) ex situ reflectance data (gray line), and reflectance calculated using the profiles in (a) (black solid line: graded profile; dotted line: optimized homogeneous profile); (c) in situ reflectance measurement (gray line), and reflectance calculated using the profiles in (a) (black solid line: graded profile; dotted line: optimized homogeneous profile).\label{exsitu}}
\end{minipage}
\end{center}
\end{figure}

For comparison additional ex situ spectrophotometric and spectroellipsometric analysis of samples from Sec.~\ref{application} have been performed.  Using the n(z) profile for SiN$_{1.3}$ layer on glass calculated with the envelope method (Fig.~\ref{exp1a}), cutting this profile in small sublayers, and introducing a small n($\lambda$) dispersion, it was found that the reflectance calculated from this profile reproduce well the ex situ measured reflectance (Fig.~\ref{exsitu}(b)) and the in situ measured reflectance (Fig.~\ref{exsitu}(c)).  In addition, a simple model considering a single homogeneous layer (dotted line in Fig.~\ref{exsitu}(a)) has been used to reproduce more precisely the ex situ reflectance (Fig.~\ref{exsitu}(b));  it was found that this model gives almost the same in situ reflectance variation with time (Fig.~\ref{exsitu}(c)) as the graded profile (solid line in Fig.~\ref{exsitu}(a)).  However, the ex situ spectra provide very little information about the exact index profile; in fact, different changes in the thickness and the n(z) values of the graded profile in Fig.~\ref{exsitu}(a) could also result in a good match with the ex situ data in Fig.~\ref{exsitu}(b).  Therefore, ex situ measurements do not offer enough reliable informations about the real profile n(z).  Ex situ spectroellipsometric measurements led to the same observations.  One can conclude that: (i) ex situ analysis can give a false impression of homogeneity, due to the fact that several different models can fit ex situ data (there are several local minima of the merit function used in the fitting procedure); (ii) in situ analysis using the envelope method proposed in this work avoids fitting problems, but small errors in the experimental data can lead to large uncertainties in the n(z) and k(z) profiles.  The advantage of in situ measurements is clear (``it's easier to investigate a murder if you can see it in situ"\cite{tompkins}), but finding n(z) very precisely requires very sensitive measurements.

\section{Conclusion}
An envelope method has been developed for the determination of refractive index profile n(z) from in situ spectrophotometric measurements (reflectance or transmittance).  The method is based on the WKBJ approximation valid for slowly varying refractive index profiles.  Generalized analytical expressions, suitable for high or low index substrates, are given for the calculation of n(z) from in situ transmittance measurements.  A more generalized solution, valid also for non-transparent substrates was calculated and it is used for the determination of n(z) from in situ reflectance measurements.  The instantaneous deposition rate and its variation during the deposition process are derived from the calculated n(z) profile.  The main feature of  the envelope method developed in this work is its analytical character: Unlike usual in situ spectrophotometric and ellipsometric analysis, no non-linear fit algorithms or ``initial guess" of the refractive index has to be used in order to ascertain n(z).  In fact, the simplicity of this envelope method makes it attractive for rapid first approximation analysis to generate initial data for more sophisticated methods such as in situ ellipsometry.

\acknowledgments
The authors are indebted to Mr. Gilles Jalbert and to Mr. Jiri Cerny for their technical assistance with the experiments.  This work was supported in part by the FCAR of Quebec and by the NSERC of Canada.


\newpage


\vglue-0.3cm

\vglue-0.3cm

\vglue-0.3cm

\vglue-0.3cm

\vglue-0.3cm

\vglue-0.3cm

\vglue-0.3cm

\vglue-0.3cm

\vglue-0.3cm

\vglue-0.3cm

\vglue-0.3cm

\end{multicols}

\end{document}